# Fano interference of the Higgs mode in cuprate high-$T_c$ superconductors

Hao Chu[1,2,3,4*†], Sergey Kovalev[5*], Zi Xiao Wang[6], Lukas Schwarz[1], Tao Dong[6], Liwen Feng[1,4,7], Rafael Haenel[1,2,3], Min-Jae Kim[1,4,7], Parmida Shabestari[1,4], Hoang Le Phuong[1,4], Kedar Honasoge[1,4], Robert David Dawson[1], Daniel Putzky[1], Gideok Kim[1], Matteo Puviani[1], Min Chen[5], Nilesh Awari[5], Alexey N. Ponomaryov[5], Igor Ilyakov[5], Martin Bluschke[1,2,3], Fabio Boschini[8], Marta Zonno[1,2,3], Sergey Zhdanovich[2,3], Mengxing Na[2,3], Georg Christiani[1], Gennady Logvenov[1], David J. Jones[2,3], Andrea Damascelli[2,3], Matteo Minola[1], Bernhard Keimer[1], Dirk Manske[1], Nanlin Wang[6,9], Jan-Christoph Deinert[5], Stefan Kaiser[1,4,7†]

[1] *Max Planck Institute for Solid State Research, Heisenbergstr. 1, 70569 Stuttgart, Germany*

[2] *Quantum Matter Institute, University of British Columbia, Vancouver, BC V6T 1Z4, Canada*

[3] *Department of Physics and Astronomy, University of British Columbia, BC V6T 1Z1, Canada*

[4] *4th Physics Institute, University of Stuttgart, 70569, Stuttgart, Germany*

[5] *Helmholtz-Zentrum Dresden-Rossendorf, Bautzner Landstr. 400, 01328 Dresden, Germany*

[6] *International Center for Quantum Materials, School of Physics, Peking University, Beijing 100871, China*

[7] *Institute of Solid State and Materials Physics, Technical University Dresden, 01062, Dresden, Germany*

[8] *Énergie Matériaux Télécommunications Research Centre, 1650 blvd Lionel-Boulet Varennes, Québec J3X 1S2, Canada*

[9] *Beijing Academy of Quantum Information Sciences, Beijing 100913, China*

[*] These authors contributed equally.
[†] Corresponding authors: haochu@mail.ubc.ca, s.kaiser@fkf.mpg.de

**Despite decades of search for the pairing boson in cuprate high-$T_c$ superconductors, its identity still remains debated to date. For this reason, spectroscopic signatures of electron-boson interactions in cuprates have always been a center of attention. For example, the kinks in the quasiparticle dispersion observed by angle-resolved photoemission spectroscopy (ARPES) studies have motivated a decade-long investigation of electron-phonon as well as electron-paramagnon interactions in cuprates[1-3]. On the other hand, the overlap between the charge-order correlations and the pseudogap in the cuprate phase diagram has also generated discussions about the potential link between them[4-6]. In the present study, we provide a fresh perspective on these intertwined interactions using the novel approach of Higgs spectroscopy, i.e. an investigation of the amplitude oscillations of the superconducting order parameter driven by a terahertz radiation. Uniquely for cuprates, we observe a Fano interference of its dynamically driven Higgs mode with another collective mode, which we reveal to be charge density wave fluctuations from an extensive doping- and magnetic field-dependent study. This finding is further corroborated by a mean field model in which we describe the microscopic mechanism underlying the interaction between the two orders. Our work demonstrates Higgs spectroscopy as a novel and powerful technique for investigating intertwined orders and microscopic processes in unconventional superconductors.**

Coupled oscillators, or the interaction between individual degrees of freedom, is a topic of fundamental importance in physics. For instance, electron-phonon interaction is at the very heart of BCS-superconductivity. In high-$T_c$ superconductors, the pairing boson is presumably different from phonons. The search for its identity has led to an extensive survey of bosonic excitations in solids and their influences on Cooper pairing[1-7]. To understand their complex interplay, past research has focused on separately examining the individual orders (e.g. superconductivity, charge order, antiferromagnetism, lattice) using different spectroscopic probes. This has led to many insightful findings including, for example, the competition between the charge order and macroscopic superconductivity[8,9], the connection between the nodal kink and particular phonon modes or magnetic excitations[1-3], and more recently a potential connection between the pseudogap and an underlying density wave order[4-6]. Nevertheless, the *dynamical* interaction between superconductivity and coexisting orders has not been visualized experimentally. Realizing such a possibility may reveal microscopic interactions behind these intertwined orders, a knowledge that is difficult to obtain from studying these orders in parts and in equilibrium.

A fresh perspective in this regard can be provided by the dynamical oscillations of the superconducting order parameter (OP) itself. The latter is equivalent to the Higgs field in high energy physics through the Anderson-Higgs mechanism, in which the gauge boson for the electromagnetic (electroweak) interaction becomes massive due to its interactions with the Cooper pairs (Higgs field)[10,11]. Therefore, the amplitude mode of the superconducting OP is also referred to as 'the Higgs mode'. It has been proposed that the Higgs mode may provide insight on the OP symmetry[12] and the pairing mechanism of a superconductor[13], making it highly relevant for studying high-$T_c$ superconductors[14,15]. In our prior investigation of cuprate high-$T_c$ superconductors, we have observed an interference of the Higgs mode with another collective mode through the characteristic anti-resonance of the former[15], in stark contrast to *s*-wave superconductors where the Higgs mode is free from such an effect[16] (SI). Qualifying the anti-resonance as a Fano resonance/interference in the present work, we show that the coupled mode can be naturally identified as charge density wave fluctuations (CDW). Our novel Higgs spectroscopy technique gives us unprecedented insights into the microscopic mechanism describing the interaction between superconductivity and CDW, an important step forward for understanding their respective roles in the cuprate phase diagram.

Figure 1 gives an overview of our experiment. As Cooper pairs themselves do not carry an electric dipole, they couple to electromagnetic radiation quadratically. Therefore, the Higgs mode can be periodically driven at $2\omega$ using a terahertz field of frequency $\omega$ (Fig. 1a)[16]. The $2\omega$ OP oscillation manifests as a modulation of optical reflectivity, which can be probed in a terahertz pump-optical probe experiment (SI). In addition, it undergoes sum frequency generation with the driving field (i.e. $2\omega + \omega$), leading to third harmonic generation (THG) in transmission (Fig. 1c). In the conventional *s*-wave superconductor NbN, the THG intensity ($I_{3\omega}$) is observed to peak at a characteristic temperature ($T$) below $T_c$, where the driving frequency $\omega$ and the superconducting OP ($2\Delta(T)$) satisfies $2\omega = 2\Delta(T)$[16] (SI). This condition arises because the resonant frequency of the Higgs mode is coincidentally also $2\Delta(T)$, therefore it becomes resonantly driven when $2\omega = 2\Delta(T)$. Here, we highlight the unique aspect of our experiment: the resonance is approached by sweeping temperature instead of the electromagnetic frequency (Fig. 1b), as is done in most spectroscopy experiments. As we increase temperature from zero, the Higgs mode is first driven below resonance (i.e. $2\omega < 2\Delta_{T=0}$) and then above (i.e. $2\omega > 2\Delta_{T\gg 0}$). As is typical for driven harmonic oscillators, right across the resonance we expect a jump in the phase of the driven Higgs oscillations (i.e. with respect to the driving field) to accompany the divergence in their amplitude[17,18]. Such a resonance-like jump is indeed observed in the THG phase ($\Phi_{3\omega}$) from NbN (SI), coinciding with the peak in $I_{3\omega}$, confirming the resonant nature of the peak. We note that additional mechanisms, such as pair-breaking as well as two- (in-plane) Josephson plasmon scattering has also been considered as sources of THG in a superconductor. In our experiment, The Higgs contribution to THG as the dominant source is distinguished holistically by several aspects of our results, as discussed in SI.

In the (optimally-doped) cuprate high-$T_c$ superconductors, we observe a similar resonance-like evolution of $\Phi_{3\omega}$ at low temperatures (Fig. 2b, see also SI and ref(15) for results from bilayer cuprates). This, however, is interrupted at some higher temperature ($T_\pi$) by an abrupt jump of nearly π in the opposite direction. The negative phase jump is also accompanied by a dip in $I_{3\omega}(T)$. Together, these features identify an anti-resonance, an essential part of the Fano resonance/interference[19,20], where our heavily damped Higgs mode and an additional underdamped collective mode play the respective roles of the continuum and discrete states (Fig. 1b). Interestingly, the Fano interference observed in our study can be characterized almost entirely by the evolution of $\Phi_{3\omega}$, which according to recent theoretical studies encodes important information about the microscopic processes within a superconductor. For example, the clean or dirty limits of a superconductor can be distinguished based on the size of the resonant phase jump (i.e. π/2 versus π)[21]. In the case of additional degrees of freedom coupled to the superconducting OP, a Fano interference may indeed show up in the Higgs response, providing important spectroscopic fingerprints for identifying and investigating intertwined interactions in a superconductor[18].

To identify the nature of the coupled mode observed in our study, first we investigate the anti-resonance as a function of hole-doping in the canonical single-layer cuprate family La$_{2-x}$Sr$_x$CuO$_4$. Figure 2 shows that the salient features of the anti-resonance can be identified at almost all doping levels from the underdoped to the overdoped regime, with $\Phi_{3\omega}(T)$ showing the most systematic evolution. The anti-resonance is found to be sharp in the underdoped and optimally doped samples and broadens with further hole-doping, making the dip in $I_{3\omega}(T)$ more visible on the overdoped side. In Fig. 2f we superpose $\Phi_{3\omega}(T)$ from all five samples on the reduced temperature ($T/T_c$) scale. In addition to $\Phi_{3\omega}(T)$ becoming increasingly broadened on the overdoped side, $T_\pi$ and $T_c$ also become increasingly separated. Together, these observations suggest a systematic change in the interaction between the Higgs mode and the coupled mode as hole-doping level increases.

According to the generalized Fano resonance model (SI), a broadening/suppression of the anti-resonance may arise from any of these scenarios: 1) the underdamped coupled mode becomes more heavily damped, 2) its frequency detunes from the heavily damped mode/continuum states, 3) the coupling is reduced. The latter two scenarios are also inter-connected as 2) can lead to 3). Directly fitting into these descriptions, particularly scenario 1), is the doping dependence of CDW fluctuations, which are expected to be well-defined on the underdoped side and heavily damped on the overdoped side. Since the interaction between CDW and superconductivity is also expected from a macroscopic persepctive[8,9], they serve as our primary candidate for the coupled mode. A generic low-frequency phonon, on the other hand, may be ruled out as hole-doping is unlikely to significantly modify its linewidth, frequency or interaction with the condensate. Last but not least, antiferromagnetic spin fluctuations, which have been extensively studied by neutron scattering, similarly become increasingly damped with hole-doping. In the following, we argue that they can be excluded based on our magnetic field dependent results as well as to the leading order on theoretical ground.

Figure 3 shows the temperature dependence of THG from the OP43 and OD30 sample in the presence of a magnetic field applied along the $c$-axis of the sample. A clear suppression of the anti-resonance is manifest in both the amplitude and phase response of THG: the dip feature in $I_{3\omega}(T)$ disappears while the sharp jump in $\Phi_{3\omega}(T)$ broadens. These results suggest that, similar to hole-doping, the external magnetic field also damps the coupled mode. To understand the implication of this, we first note that the low- and high-energy spin fluctuations show distinctively different response to a $c$-axis magnetic field. The low energy spin fluctuations, which are gapped out by the superconducting transition, can be re-introduced at low energies by a $c$-axis magnetic field[22]. Within the Fano model, this would not lead to a broadening of the phase jump as we observed experimentally. On the other hand, the high energy spin fluctuations (a.k.a. the magnetic resonance mode), which take spectral weight from the low energy part during the superconducting transition, respond weakly to a *c*-axis field[23,24]. Furthermore, their energy resides far above our terahertz driving frequency, making the anti-resonance inaccessible in our experiment even if such a coupling exists (Fig. 1b). Based on these experimental facts, we conclude that both the low- and high-energy spin fluctuations are unlikely to explain the magnetic field dependence of the anti-resonance.

Having ruled out antiferromagnetic spin fluctuations as the coupled mode, next we consider CDW amplitude fluctuations, which are understood as mediated by phonons at the CDW ordering wavevector $q_{cdw}$ in most systems where either Fermi surface nesting or anisotropic electron-phonon coupling (EPC) drives the CDW instability[25]. Their experimental signature, i.e. a modulation of the CDW gap at such phonon frequencies, has been identified on layered chalcogenides by time-resolved ARPES studies[26]. Additional optical signature, i.e. a modulation of transient reflectivity at these phonon frequencies, has also been identified by time-resolved optical experiments[27]. The diverging CDW correlation at $T_{CDW}$ in these systems leads to a softening of the $q_{cdw}$ phonons[27]. This has been similarly observed for underdoped cuprates, albeit at a temperature markedly lower than $T_{CDW}$[28], indicating that the true long-range ordering of CDW in cuprates might be significantly below $T_{CDW}$. In optimally and overdoped cuprates, the CDW correlation is noticeably weaker compared to the underdoped side[29], therefore phonons do not exhibit visible softening at $q_{cdw}$ at low temperatures as confirmed also by experiment[28]. However, as we apply a magnetic field, the superconducting long-range order is suppressed while the CDW correlation is enhanced[8,9]. It is then reasonable to expect the latter to move closer towards the critical point and induce a (partial) softening of the $q_{cdw}$ phonons at low temperatures. Consequently, the CDW fluctuations mediated by these phonons also soften, a scenario that is consistent with the field-induced broadening of the anti-resonance in Fig. 3. We note that the partial softening of the $q_{cdw}$ phonons can also be effected by a uniaxial pressure[30].

Our discussion above relies on the assumption that the CDW fluctuations in our study arise from a phonon-mediated mechanism. It does not presume nesting or anisotropic EPC as the main driving force behind the CDW instability in cuprates. Interestingly, recent theoretical studies have highlighted the importance of incipient softening of phonons, in conjunction with electronic correlation, in driving the peculiar formation of CDW in cuprates[31]. Experimentally, a phonon-like modulation of transient reflectivity, whose amplitude appears to track the CDW OP as a function of temperature, has also been reported and interpreted as the amplitude mode of CDW in cuprates and related compound[32-34]. In particular, a coupling between the superconducting OP and the CDW OP has already been considered phenomenologically in ref(33). Taking a step further, we consider a microscopic mean field model previously developed for the Raman response of coexisting superconducting and CDW state in NbSe$_2$[35,36]. We use this model to illustrate, in the most general sense, the interference between the Higgs mode and phonon-mediated CDW amplitude fluctuations in a THG experiment (SI). As shown in Fig. 4, our model predicts a single resonance in THG if the Higgs mode is independent from other collective modes, similar to the results from NbN. In comparison, if the Higgs mode couples to the CDW fluctuations, an additional anti-resonance appears, in qualitative agreement with our findings from cuprates. A similar investigation of the Higgs-spin fluctuations interaction suggests that the latter is much weaker compared to the Higgs-CDW interaction. Higher order processes, such as 2-magnon scattering[37], might need to be considered to see if a significant interaction with the Higgs mode is possible. In the present study, we conclude that the phonon-mediated CDW fluctuations provide the best explanation for the coupled mode. The requirement of having *two* coupled modes in order to produce the Fano interference in a THG experiment also helps distinguish the Higgs mode, among all proposed collective modes or mechanisms, as the essential contributor to THG in our experiment (SI).

We note also another interesting finding from the magnetic field dependence results, which may elucidate the microscopic length scale of the THG process. Despite significant changes in $I_{3\omega}(T)$ near $T_\pi$, we notice that the overall suppression of THG by the applied field is small up to 10 T in both samples above and below $T_c$. This can be clearly seen for the OD30 sample in Fig. 3c but is more evident for both samples from the field-sweep results performed at constant temperatures (see SI). As the nonlinear signal is sensitive to small fluctuations and drifts in the driving power (i.e. $I_{3\omega} \propto I_\omega^3$) and a complete temperature-sweep as shown in Fig. 3 takes many hours to acquire, the true field dependence of THG is more accurately reflected by the field-sweep results. The lack of field dependence in $I_{3\omega}$ away from $T_\pi$ suggests that the length scale relevant to the THG process is more likely related to the superconducting coherence length ($\xi_{ab}$ ~ 20-30 Å in cuprates) than the London penetration depth ($\lambda_{ab}$ ~ several hundred nm). The latter is known to increase in the presence of a magnetic field[38], which would have led to a decrease in $I_{3\omega}$ at any given temperature. The coherence length, on the other hand, is more robust to external fields as it scales with the upper critical field $H_{c2}$ as $\xi_{ab} \propto H_{c2}^{-1/2}$, where $H_{c2}$ is on the order of several tens of Teslas in these materials. Adopting such an interpretation would immediately suggest that the above-$T_c$ THG observed in our studies[15] may arise from local preformed Cooper pairs. In comparison, in NbN THG drops to zero above $T_c$, consistent with an absence of preformed Cooper pairs in conventional superconductors[16] (SI).

We note that in some of our measurements, an above-$T_c$ anti-resonance phase jump may also be inferred (SI), giving additional support for the picture of preformed Coopers and a nonzero interaction between them and the incipient CDW above $T_c$. Such an interpretation naturally reminds the intimate relation between the two OP's in the pseudogap region of the cuprate phase diagram. A microscopic and in-depth investigation of their interactions is now readily available, as demonstrated by our novel phase-resolved Higgs spectroscopy technique. We note that a recent theoretical model, i.e. the pair density wave (PDW) model, predicts an interaction between the two OP's already above $T_c$ as they onset from the same PDW parent state[39]. To investigate the potential link between the PDW model and our findings, and to unlock the longstanding mysteries of high-$T_c$ superconductors, we anticipate future studies to extend Higgs spectroscopy to all possible directions.

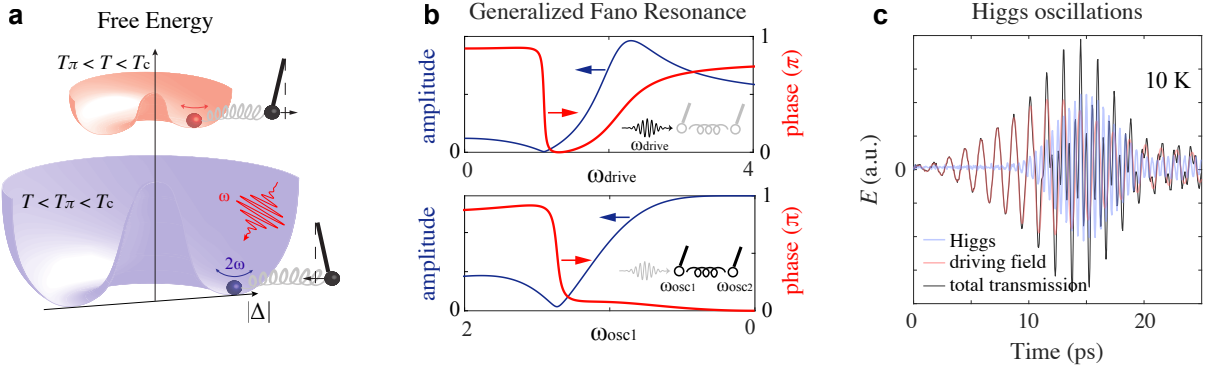

**Figure 1 Higgs spectroscopy of *d*-wave superconductors. a,** An illustration of the amplitude oscillation of the superconducting order parameter (2ω), i.e. the Higgs mode, driven by electromagnetic radiation (ω) and its coupling to an additional collective mode (black pendulum) at two temperatures. $T_\pi$ is the anti-resonance temperature. **b,** The generalized Fano resonance model describes the interference between a driven damped (continuum) mode and an underdamped (discrete) mode, here represented as oscillator 1 and 2 respectively. The Fano interference/resonance is characterized by the asymmetrical line-shape in the amplitude response, and also by the negative π jump in the phase response (a.k.a. anti-resonance). Upper: amplitude and phase response of oscillator 1 as the driving frequency $\omega_{drive}$ is swept. Lower: keeping the driving frequency fixed and sweeping instead the resonance frequency of oscillator 1 (i.e. analogous to sweeping temperature for a temperature-dependent collective mode). Notice the direction of the horizontal axis: $\omega_{osc1}$ increases to the left so that $\omega_{osc1} > \omega_{drive}$ on the left side of the figure and vice versa, in line with the two limits of the upper figure (SI). **c,** Terahertz transmission from a superconducting LSCO ($T_c$ = 44 K) thin film as it is pumped by a 0.7 THz driving pulse. The latter drives the 2ω Higgs oscillation (see SI for a clear experimental visualization of this 2ω oscillation from reflectivity measurements), which then undergoes sum frequency generation with the driving field to produce third harmonic generation (THG) in the transmission experiment. The driving field (red) and the THG (blue) waveforms are extracted from the raw transmission data (black) using 1.4 THz Fourier lowpass and highpass filters. From their respective waveforms, the relative phase between THG and the driving field can be extracted (SI).

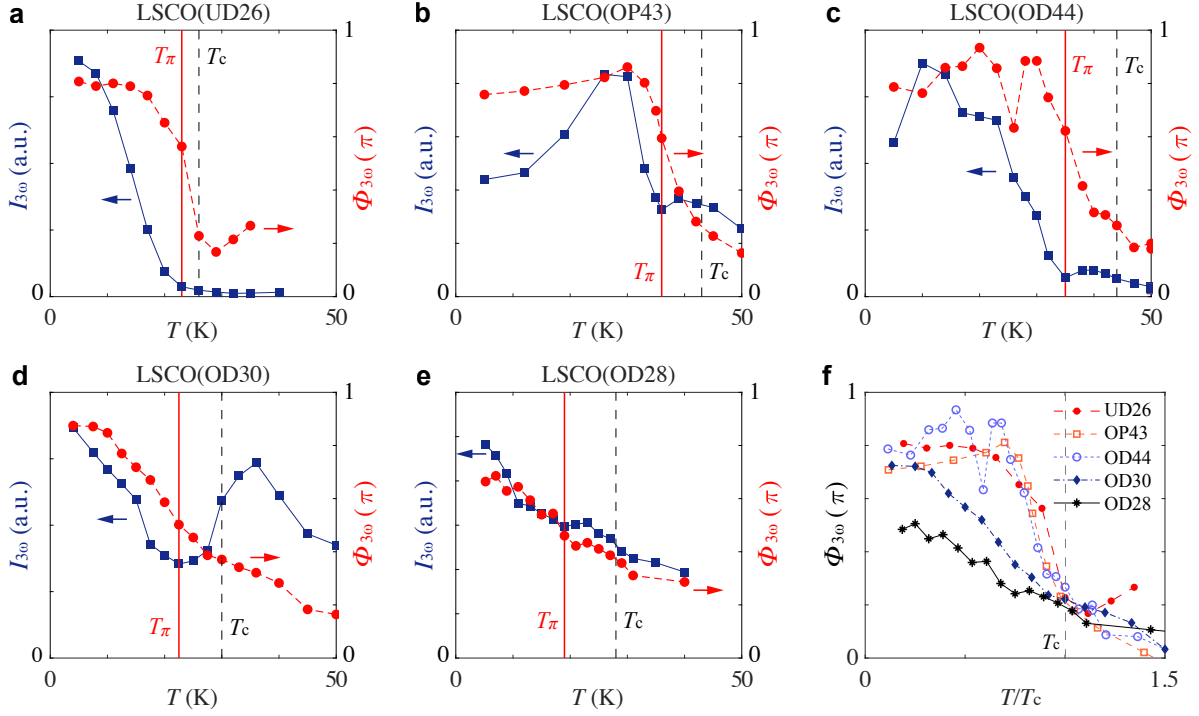

**Figure 2 Doping dependence of Fano interference in $La_{2-x}Sr_xCuO_4$.** The Fano interference of the driven Higgs oscillations is identified by the dip in THG intensity ($I_{3\omega}$) concomitant with the negative jump in THG phase ($\Phi_{3\omega}$) at a temperature denoted by $T_\pi$ (red solid line). The dip-peak feature in $I_{3\omega}(T)$ near and above $T_\pi$ gives the typical asymmetrical line shape associated with the Fano resonance/interference, compounded by an additional superconducting screening factor[15] (i.e. the electromagnetic driving field becomes increasingly screened at lower temperature by the superconducting transition. In other words, the driving force in the generalized Fano resonance model becomes smaller as temperature decreases. This leads to, in some cases, a peak in $I_{3\omega}(T)$ below $T_\pi$, such as in **b**, which is not associated with the resonance of the Higgs mode, as confirmed by the absence of a resonance phase jump in $\Phi_{3\omega}$ around that temperature). **a-e,** Temperature dependence of $I_{3\omega}$ (blue squares) and $\Phi_{3\omega}$ (red dots) in **(a)** UD26 ($x \sim 0.12$), **(b)** OP43 ($x \sim 0.16$), **(c)** OD44 ($x \sim 0.25$), **(d)** OD30 ($x \sim 0.30$), **(e)** OD28 ($x \sim 0.35$). **(f)** Temperature dependence of $\Phi_{3\omega}$ from all five samples plotted on the reduced temperature ($T/T_c$) scale. The anti-resonance becomes increasingly broadened with hole-doping, indicating a broadening of the coupled mode on the overdoped side.

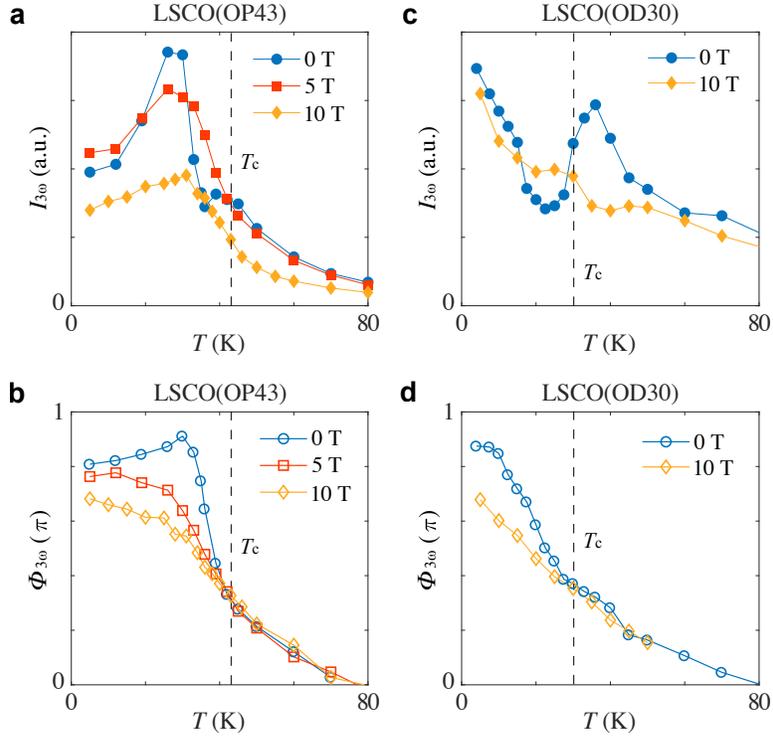

**Figure 3 Magnetic field dependence of Fano interference in $La_{2-x}Sr_xCuO_4$. a, c,** Temperature dependence of $I_{3\omega}$ with a magnetic field applied along the *c*-axis of **(a)** OP43, **(c)** OD30. **b, d,** Corresponding temperature dependence of $\Phi_{3\omega}$ from the two samples. The anti-resonance becomes increasingly broadened by the magnetic field, indicating a broadening/softening of the coupled mode with magnetic field.

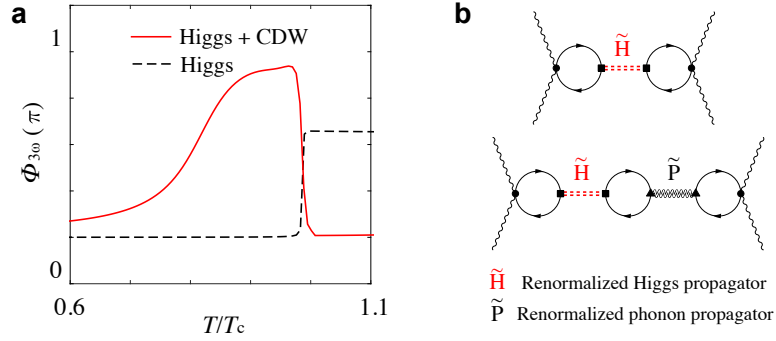

**Figure 4 Mean field model for the Higgs mode-CDW fluctuations interaction. a,** Temperature dependence of $\Phi_{3\omega}$ from an independent Higgs mode (black dashed line), and a CDW-coupled Higgs mode (red solid line) under terahertz drive, predicted from a mean field model described in detail in (SI). The former exhibits a single resonance with a positive $\pi/2$ phase jump. The latter exhibits (in addition to resonances) an anti-resonance with a negative $\pi$ phase jump. **b,** Feynman diagrams for the two THG processes shown in **a**. Light scattering off (top) an independent Higgs mode, (bottom) a CDW-coupled Higgs mode, where the CDW fluctuations are mediated by phonons.

**Reference**

1. Bogdanov, P. *et al.* Evidence for an Energy Scale for Quasiparticle Dispersion in $Bi_2Sr_2CaCu_2O_8$. *Phys. Rev. Lett.* **85** 2581 (2000).
2. Lanzara, A. *et al.* Evidence for ubiquitous electron-phonon coupling in high-$T_c$ superconductor. *Nature* **412**, 510 (2001).
3. Eschrig, M. The effect of collective spin-1 excitations on electronic spectra in high-$T_c$ superconductors. *Advances in Physics* **55**:1-2, 47-183 (2006).
4. Hashimoto, M. *et al.* Particle–hole symmetry breaking in the pseudogap state of Bi2201. *Nat. Phys.* **6**, 414-418 (2010).
5. Comin, R. *et al.* Charge Order Driven by Fermi-Arc Instability in $Bi_2Sr_{2-x}La_xCuO_{6+d}$. *Science* **343**, 390-392(2014).
6. Lee, P. A. Amperean Pairing and the Pseudogap Phase of Cuprate Superconductors. *Phys. Rev. X* **4**, 031017 (2014).
7. Keimer, B. *et al.* From quantum matter to high-temperature superconductivity in copper oxides. Nature **518**, 179-186 (2015).
8. Ghiringhelli, G. *et al.*, Long-Range Incommensurate Charge Fluctuations in $(Y,Nd)BaCuO_{6+x}$. Science **337**, 821-825 (2012).
9. Chang, J. *et al.* Direct observation of competition between SC and CDW order in $YBa_2Cu_3O_{6.67}$. *Nat. Phys.* **8**, 871 (2012).
10. Anderson, P. W. Plasmons, Gauge Invariance, and Mass. *Phys. Rev.* **130**, 439-442 (1963).
11. Higgs, P. W. Broken Symmetries and the Masses of Gauge Bosons. *Phys. Rev. Lett.* **13**, 508-509 (1964).
12. Schwarz, L. *et al.* Classification and characterization of nonequilibrium Higgs modes in unconventional superconductors. *Nat. Commun.* **11**, 287 (2020).
13. Murakami, Y. *et al.* Multiple amplitude modes in strongly coupled phonon-mediated superconductors. *Phys. Rev. B* **93**, 094509 (2016).
14. Katsumi, K. *et al.* Higgs Mode in the d-Wave Superconductor Bi2Sr2CaCu2O8+x Driven by an Intense Terahertz Pulse. *Phys. Rev. Lett.* **120**, 117001 (2018).
15. Chu, H. *et al.* Phase-resolved Higgs response in superconducting cuprates. *Nat. Commun.* **11** 1793 (2020).

16. Matsunaga, R. *et al.* Light-induced collective pseudospin precession resonating with Higgs mode in a superconductor. *Science* **345**, 1145-1149 (2014).
17. Tsuji, N. & Aoki, H. Theory of Anderson pseudospin resonance with Higgs mode in superconductors. *Phys. Rev. B* **92**, 064508 (2015).
18. Schwarz, L. *et al.* Phase signatures in third-harmonic response of THz driven superconductors. https://arxiv.org/abs/2107.01445
19. Limonov, M. F. *et al.* Fano Resonances in photonics. *Nat. Photon.* **11**, 543-554 (2017).
20. Luk'yanchuk, B. *et al.* The Fano resonance in plasmonic nanostructures and metamaterials. *Nat. Mat.* **9**, 707-715 (2010).
21. Haenel, R. *et al.* Time-resolved optical conductivity and Higgs oscillations in two-band dirty superconductors. https://arxiv.org/abs/2012.07674
22. Lake, B. *et al.* Spins in the Vortices of a High-Temperature Superconductor. *Science* **291**, 1759-1762 (2001).
23. Dai, P. *et al.*, Resonance as a measure of pairing correlations in the high-$T_c$ superconductor $YBa_2Cu_3O_{6.6}$. *Nature* **406**, 965 (2000).
24. Fujita, M. *et al.* Progress in Neutron Scattering Studies of Spin Excitations in High-$T_c$ Cuprates. *J. Phys. Soc. Jpn.* **81**, 011007 (2012).
25. Gruner, G. The dynamics of charge-density waves. *Rev. Mod. Phys.* **60**, 1129-1181 (1988).
26. Schmitt, F. *et al.* Transient Electronic Structure and Melting of a Charge Density Wave in $TbTe_3$. *Science* **321**, 1649-1652 (2008).
27. Schafer, H. *et al.* Disentanglement of the Electronic and Lattice Parts of the Order Parameter in a 1D Charge Density Wave System Probed by Femtosecond Spectroscopy. *Phys. Rev. Lett.* **105**, 066402 (2010).
28. Tacon, M. L. *et al.* Inelastic X-ray scattering in $YBa_2Cu_3O_{6.6}$ reveals giant phonon anomalies and elastic central peak due to charge-density-wave formation. *Nat. Phys.* **10**, 52-58 (2014).
29. Wen, J. J. *et al.* Observation of two types of charge-density-wave orders in superconducting $La_{2-x}Sr_xCuO_4$. *Nat. Commun.* **10**, 3269 (2019).
30. Kim, H.-H. *et al.* Uniaxial pressure control of competing orders in a high-temperature superconductor. *Science* **362**, 1040 (2018).
31. Banerjee, S. *et al.* Emergent charge order from correlated electron-phonon physics in cuprates. *Commun. Phys.* **3**, 161 (2020).

32. Torchinsky, D. *et al.* Fluctuating charge-density waves in a cuprate superconductor. *Nat. Mat.* **12**, 387-391 (2013).

33. Hinton, J. P. *et al.* New collective mode in $YBa_2Cu_3O_{6+x}$ observed by time-domain reflectometry. *Phys. Rev. B* **88**, 060508(R) (2013).

34. Chu, H. *et al.* A charge density wave-like instability in a doped spin–orbit-assisted weak Mott insulator. *Nat. Mat.* **16**, 200-203 (2017).

35. Méasson, M.-A. *et al.* Amplitude Higgs mode in the $2H-NbSe_2$ superconductor. *Phys. Rev. B* **89**, 060503 (2014).

36. Cea, T. & Benfatto, L. Nature and Raman signatures of the Higgs amplitude mode in the coexisting superconducting and charge-density-wave state. *Phys. Rev. B* **90**, 224515 (2014).

37. Venturini, F. *et al.* Collective Spin Fluctuation Mode and Raman Scattering in Superconducting Cuprates. *Phys. Rev. B* **62**, 15204 (2000).

38. Schafgans A. A. *et al.* Towards a Two-Dimensional Superconducting State of $La_{2-x}Sr_xCuO_4$ in a Moderate External Magnetic Field. *Phys. Rev. Lett.* **104**, 157002 (2010).

39. Agterberg, D. F. *et al.* The Physics of Pair-Density Waves: Cuprate Superconductors and Beyond. *Annu. Rev. Condens. Matter Phys.* **11**, 231-270 (2020).

**Acknowledgements**

The authors thank Jingda Wu, Ziliang Ye, Francesco Gabriele, Lara Benfatto for valuable discussions, and the ELBE team for the operation of the TELBE facility. Parts of this research were carried out at ELBE at the Helmholtz-Zentrum Dresden - Rossendorf e. V., a member of the Helmholtz Association. This research was undertaken also thanks in part to funding from the Max Planck-UBC-UTokyo Centre for Quantum Materials and the Canada First Research Excellence Fund, Quantum Materials and Future Technologies Program. This project is also funded by the Natural Sciences and Engineering Research Council of Canada (NSERC); the Alexander von Humboldt Fellowship (A.D.); the Canada Research Chairs Program (A.D.); and the CIFAR Quantum Materials Program.

**Author contributions**

H.C. and S.Ka. conceived the project. H.C., S.Ka., S.Ko., J.-C.D. developed the experimental plan. Measurements on the cuprate samples were performed at the TELBE facility. S.Ko., H.C., L.F., M.-J.K., P.S., H.L.P., K.H., R.D.D., M.C., N.A., A.N.P., I.I., M.Z., S.Z., M.N., J.-C.D., S.Ka. conducted the beamtime experiment. Measurements on NbN were performed by Z.X.W., T.D., N.W. at Peking University. H.C. performed data analysis and interpreted the results together with M.B., F.B., M.M., D.M., S.Ka. and B.K.. L.S., R.H., M.P., D.M. performed theoretical modeling. G.K., D.P., G.C., G.L. provided and characterized the cuprate samples. H.C. wrote the manuscript with S.Ka., D.M., A.D., L.S., M.P., with inputs from all authors.

**Competing interests**

The authors declare that they have no competing interests.

**Data Availability**

The data that support the findings of this study are available from the corresponding authors upon reasonable request. For raw pre-binned data that allow statistical analysis, request should be sent to HZDR via S. Ko.

# Supplementary Information

S1. Experimental method

S2. Sample characterization

S3. Terahertz pump optical reflectivity probe results

S4. THG from NbN

S5. Other proposed mechanisms of THG inside a superconductor

S6. Generalized Fano resonance model

S7. Magnetic field-sweep results

S8. Microscopic model for the interplay between superconductivity and CDW

S9. Evidence for anti-resonance above $T_c$

## S1. Experimental method

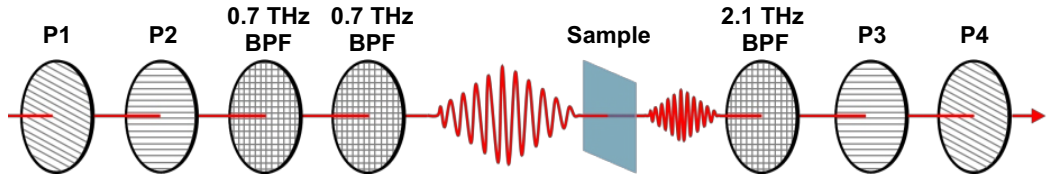

**Figure S1 Experimental setup for measuring THz third harmonic generation.** P1-P4: wire-grid polarizers. BPF: bandpass filter.

The THG experiment is performed using the schematic shown in Fig. S1. The multicycle terahertz driving pulse is generated from the TELBE super-radiant undulator source at HZDR[1]. For the data presented in the main text, we use a driving frequency of 0.7 THz and place a 2.1 THz bandpass filter after the sample for filtering the THG response. The transmitted terahertz pulse is measured using electro-optical sampling inside a 2 mm ZnTe crystal and by using 100 fs gating pulses with a central wavelength of 800 nm. The accelerator-based driving pulse and the laser gating pulse have a timing jitter characterized by a standard deviation of ~ 20 fs. Synchronization is achieved through pulse-resolved detection as detailed in ref(2).

To extract the relative phase between the THG response and the linear driving field, we fit a Gaussian-enveloped sinusoidal function, as detailed in ref(3), to both the transmitted linear driving field waveform and the THG waveform. From these fits, we obtain the phase of the driving field and the THG response for each temperature. We then take their difference (modulo $2\pi$) while accounting for the fact that a $2\pi$ phase change for the driving field corresponds to a $6\pi$ phase change for the THG response. The result is double-checked by another method as detailed in the supplementary information of ref(3), where we first manually shift the transmitted driving field in time to overlap them in phase (i.e. getting rid of the temperature-dependent phase shift induced by the superconducting screening effect), and then apply the same time shifts to the THG waveforms at the corresponding temperatures. Then we extract the THG phase by fitting a Gaussian-enveloped sinusoidal function to the time-shifted THG waveforms. Both analysis methods yield very similar results, confirming the validity of the anti-resonance phase jump in the THG response.

## S2. Sample characterization

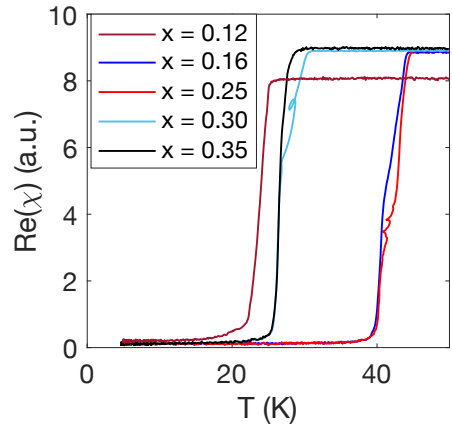

**Figure S2 Magnetic susceptibility measurements of $T_c$ on $La_{2-x}Sr_xCuO_4$** The superconducting transition temperature $T_c$ is determined from the real part of the magnetic susceptibility $\chi$. $T_c$ is defined as the temperature for the initial drop in Re($\chi$).

The $La_{2-x}Sr_xCuO_4$ samples are grown by molecular beam epitaxy (MBE) method on $LaSrAlO_4$ substrate. All samples are 40 nm thick. The $DyBa_2Cu_2O_{7-x}$ samples (see **S9**) are also grown by MBE method on $(LaAlO_3)_{0.3}(Sr_2TaAlO_6)_{0.7}$ (LSAT) substrate. The OP 90 sample is 70 nm thick. The UD50 sample is 20 nm thick. $T_c$ is determined from mutual inductance measurements as shown in Fig. S2 for the $La_{2-x}Sr_xCuO_4$ series.

## S3. Terahertz pump optical reflectivity probe results

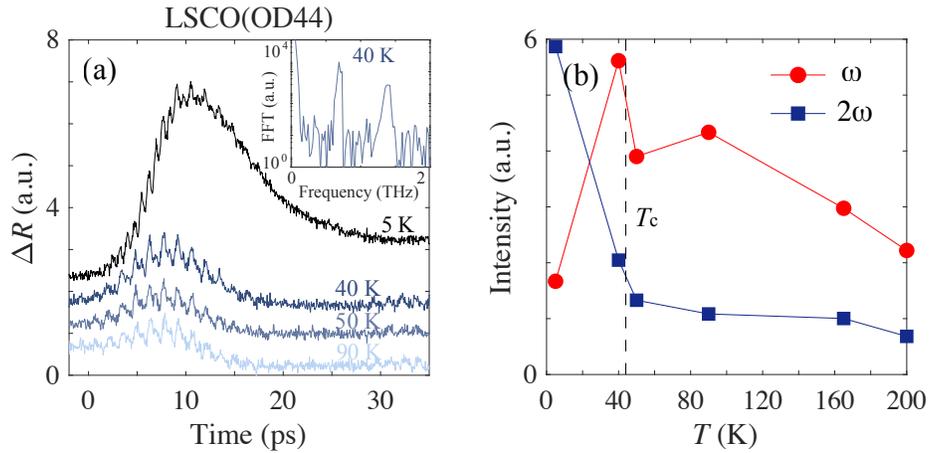

**Figure S3  Terahertz-driven 2ω Higgs oscillations in transient reflectivity**  **(a)** Transient reflectivity at 800 nm wavelength measured as a function of terahertz pump-optical probe delay from an slightly overdoped LSCO thin film. The terahertz driving field is centered around 0.7 THz. The driven Higgs oscillations at 1.4 THz is manifest in both the time-domain reflectivity and its FFT (inset).  **(b)** Temperature dependence of the intensity of the linear (ω) and Higgs (2ω) components, as extracted from FFT.

The driven Higgs oscillations can be unambiguously identified in a terahertz pump optical reflectivity probe (TPOP) measurement, in which a linear driving field at frequency ω induces the superconducting order parameter to oscillate at 2ω. Spectroscopic fingerprints of this 2ω oscillation can be obtained from optical reflectivity measurement, as is done for an LSCO(OD44) thin film shown above. Fig. S3a shows that while the 0.7 THz driving pulse is on, optical reflectivity of the sample exhibits a clear 1.4 THz modulation on top of the 0.7 THz modulation. Interestingly, while this 2ω response enhances dramatically below $T_c$, it remains non-vanishing above $T_c$ up to 200 K (Fig. S3b). The ω response, on the other hand, diverges near $T_c$, and could be associated with the pair-breaking process. The terahertz-driven current that gives rise to the ω response becomes screened below $T_c$ by the superconducting transition, consistent with the drop of the ω amplitude at lower temperature. The presence of a sizeable Higgs response in our TPOP experiment is also in agreement with the experimental results of ref(4) on BSCCO single crystals.

## S4. THG from NbN

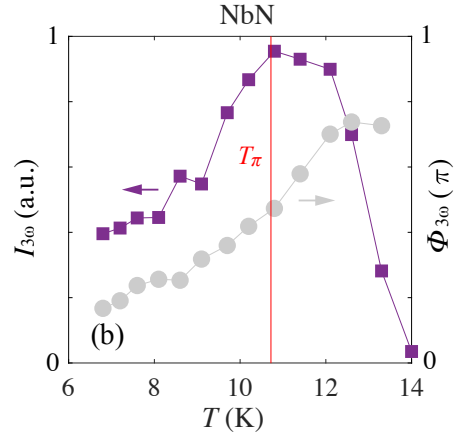

**Figure S4  THG from NbN**  The temperature dependence of THG intensity ($I_{3\omega}$, violet squares) and phase ($\Phi_{3\omega}$, grey dots) from a NbN thin film.

THG from a NbN thin film ($T_c$ = 14 K) is measured with a 0.5 THz driving field from a table-top terahertz source. As shown in Fig. S4, the THG phase ($\Phi_{3\omega}$) exhibits a resonance-like evolution with temperature (i.e. a positive phase jump). The middle of the phase jump coincides with the peak in THG intensity ($I_{3\omega}$) at the temperature $T_\pi$. In addition, THG drops to zero above $T_c$. Therefore, the THG measurement terminates right at $T_c$.

## S5. Other proposed mechanisms of THG inside a superconductor

In addition to THG arising from driven Higgs oscillations as discussed in the main text[5-8], several additional mechanisms for THG has been considered in the context of a superconductor, including the pair-breaking/BCS process[9,10], and light scattering off two Josephson plasmons[11]. Regarding the relative intensity of the Higgs and BCS contributions to THG, theory suggests that the BCS could dominate in clean *s*-wave superconductors[12], while the Higgs contribution still prevails in the dirty limit[10,12-15]. However, for *d*-wave cuprates superconductors, recent experimental and theoretical works[4,16] suggest that additional mechanisms can be responsible for a prevailing Higgs response, even without impurities.

The different mechanisms considered in these studies also lead to different polarization dependence in THG, which, depending on the level of disorder in the system, may be more or less pronounced[10,11,17]. THG originating from the BCS process as well as the two-plasmon process is expected to exhibit an anisotropic polarization dependence in agreement with the symmetry of the underlying electronic structure, and is allowed in both parallel and cross-polarized channels (as an example, see ref(18) for optical THG anisotropy from $Sr_2IrO_4$, a compound with very similar symmetry and electronic structure to $La_2CuO_4$). In contrast, THG originating from the Higgs mode in clean superconductors is expected to exhibit an isotropic polarization dependence, and is polarized parallel to the incoming driving field. As shown in our previous study[3], THG from cuprate thin films is characterized by a dominant isotropic component, with a small anisotropic component potentially on top. In the present study, we have also verified that THG in the cross-polarized channel is at least an order of magnitude weaker compared to the parallel-polarized channel. Together, these results are consistent with a dominant Higgs contribution to THG in our experiment. The great enhancement of THG below $T_c$ as a signature of the growing Higgs response (due to the growing superconducting OP) is also testified by the pronounced 2ω component in the terahertz pump optical reflectivity probe (TPOP) measurement (**S3**) below $T_c$. In addition, the persistence of the 2ω component above $T_c$ in these TPOP measurements is also in line with the non-vanishing THG above $T_c$, suggesting a common physical origin of the two types of nonequilibrium signals, namely the driven Higgs oscillations.

An additional piece of evidence for the dominant Higgs contribution to THG in our experiment comes from the observation of the Fano resonance/interference itself. As our mean field model (**S8**) indicates, light scattering off an independent collective mode (e.g. the Higgs mode, or the CDW amplitude mode) gives rise to a single resonance in THG accompanied by a positive phase jump. In order to produce the experimentally observed anti-resonance/Fano interference, theory requires two collective modes to interact with each other. Among the proposed THG mechanisms as discussed above, this requires an interaction between the Higgs mode, the Josephson plasmon, and the CDW amplitude mode as considered in this work. Based on the present understanding of cuprates, an interaction between the Higgs mode and the CDW is the most likely scenario among all the potential interactions between these collective modes. This then suggests that the main microscopic processes behind THG, at least in our studies, are described by the Feynman diagrams as shown in Fig. 4b and Fig. S6g of this manuscript, where the Higgs mode is a key ingredient.

## S6. Generalized Fano resonance model

In our previous study[3], we have identified the anti-resonance as a result of coupling to an additional collective mode based on a driven coupled harmonic oscillators model. This model provides the basis for the generalized Fano resonance model: the well-known formula of the Fano resonance is an approximation of this model near the coupled discrete/underdamped mode[19]. In the following, we recapitulate the essence of ref(19) and show that the two models are essentially equivalent.

The equations of motion of a coupled oscillators system in which one is driven periodically are given by

$$\ddot{x}_1 + \gamma_1 \dot{x}_1 + \omega_1^2 x_1 + \omega_1^2 x_1 + v_{12} x_2 = F_0 \cos(\omega t),$$

$$\ddot{x}_2 + \gamma_2 \dot{x}_2 + \omega_2^2 x_2 + v_{12} x_2 = 0,$$

where $\omega_1$, $\gamma_1$, $\omega_2$, $\gamma_2$ are the frequency and damping constant of oscillators 1 and 2 respectively. $v_{12}$ is the coupling constant, $F_0$ is the driving force. $x_1$ is found to be

$$x_1 = \frac{\omega_2^2 - \omega^2 + i\gamma_2\omega}{(\omega_1^2 - \omega^2 + i\gamma_1\omega)(\omega_2^2 - \omega^2 + i\gamma_2\omega) - v_{12}^2} F_0 \cos(\omega t)$$

Alternatively, we can find the new eigenmodes of the coupled system as

$$\omega_a^2 = \omega_1^2 - \frac{v_{12}^2}{\omega_2^2 - \omega_1^2},$$

$$\omega_b^2 = \omega_2^2 + \frac{v_{12}^2}{\omega_2^2 - \omega_1^2}.$$

Defining $\tilde{\epsilon} \equiv \omega^2 - \omega_b^2 = (\omega + \omega_b)(\omega - \omega_b)$, which is proportional to $(\omega - \omega_b)$ in the neighborhood of $\tilde{\epsilon} \sim 0$, i.e. $\omega \sim \omega_b$, we can define the reduced energy $\epsilon$ for the system

$$\epsilon \equiv \frac{1}{\gamma_1 \omega_2} \frac{(\omega_2^2 - \omega_1^2)^2}{v_{12}^2} \tilde{\epsilon}$$

Assume $\gamma_2 = 0$, we can transform the numerator of $x_1$ near $\omega_b$ (i.e. $\omega \sim \omega_b$ or $\tilde{\epsilon} \sim 0$) as

$$\omega_2^2 - \omega^2 = -\tilde{\epsilon} - \frac{v_{12}^2}{\omega_2^2 - \omega_1^2}$$

and the denominator as

$$(\omega_1^2 - \omega^2 + i\gamma_1\omega)(\omega_2^2 - \omega^2) - v_{12}^2$$

$$= (\omega_2^2 - \omega_1^2)\tilde{\epsilon} + (\omega^2 - \omega_2^2 - i\gamma_1\omega)\tilde{\epsilon} + (\omega^2 - \omega_2^2 - i\gamma_1\omega)\frac{v_{12}^2}{\omega_2^2 - \omega_1^2}$$

$$\approx (\omega_2^2 - \omega_1^2)\tilde{\epsilon} - i\gamma_1\omega \frac{v_{12}^2}{\omega_2^2 - \omega_1^2}$$

Plugging these expressions back to $x_1$, using the definition of $\epsilon$ above we have

$$x_1 = \frac{-\tilde{\epsilon} - \frac{v_{12}^2}{\omega_2^2 - \omega_1^2}}{(\omega_2^2 - \omega_1^2)\tilde{\epsilon} - i\gamma_1\omega \frac{v_{12}^2}{\omega_2^2 - \omega_1^2}} F_0 \cos(\omega t) = \frac{\gamma_1\omega_2 \frac{v_{12}^2}{(\omega_2^2 - \omega_1^2)^2}\left(-\epsilon - \frac{\omega_2^2 - \omega_1^2}{\gamma_1\omega_2}\right)}{\gamma_1\omega_2 \frac{v_{12}^2}{(\omega_2^2 - \omega_1^2)^2}(\epsilon - i)} \frac{F_0 \cos(\omega t)}{\omega_2^2 - \omega_1^2}$$

$$= -\frac{(\epsilon + q)}{(\epsilon - i)} \frac{F_0 \cos(\omega t)}{\omega_2^2 - \omega_1^2}, \qquad \text{where } q \equiv \frac{\omega_2^2 - \omega_1^2}{\gamma_1\omega_2}.$$

Here we see that the amplitude square (i.e. intensity) of $x_1$ becomes

$$|x_1|^2 \sim \frac{(\epsilon + q)^2}{\epsilon^2 + 1} \frac{F_0^2}{(\omega_2^2 - \omega_1^2)^2} \propto \frac{(\epsilon + q)^2}{\epsilon^2 + 1}$$

recovering the familiar Fano resonance formula.

For plotting Fig. 1b of the main text, we have used the following parameters and assumptions. For the upper figure, we used $\omega_1 = 1$, $\omega_2 = 1.1$, $\gamma_1 = 0.2$, $\gamma_2 = 0.02$ (introduced here to give a finite width to the anti-resonance), $v_{12} = 0.7$ while sweeping the driving frequency $\omega$. For the lower figure, $\omega = 1.5$ (fixed driving frequency), $v_{12} = 0.7$. $\omega_1$ is swept while keeping the damping constant as a fixed ratio $\gamma_1 = 0.2\omega_1$. $\omega_2$ and $\gamma_2$ are also kept as a fixed ratio of $\omega_1$: $\omega_2 = 1.1\omega_1$ and $\gamma_2 = 0.02\omega_1$. That is, the ratio between $\omega_1$, $\gamma_1$, $\omega_2$, $\gamma_2$ are identical across the upper and lower figures.

## S7. Magnetic field-sweep results

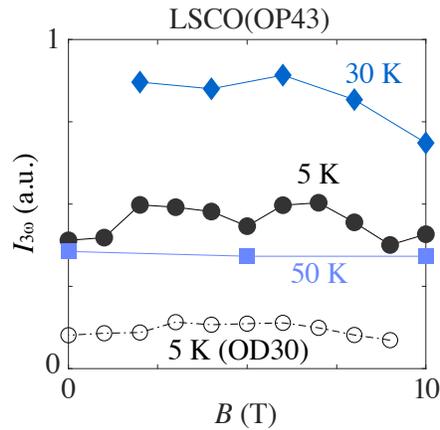

**Figure S5  THG intensity as a function of magnetic field at constant temperatures**

We measured the THG response of the OP43 and OD30 sample as a function of magnetic field (along *c*-axis) at constant temperatures. It can be seen that away from $T_\pi$ (~ 36 K for OP43 and ~ 22 K for OD30), the THG intensity changes very weakly with magnetic field up to 10 Tesla. Each data point shown here takes about 15 minutes to acquire, therefore one field-sweep can be done within one to two hours. Therefore, this measurement is less prone to fluctuations or drifts in the driving pulse power, as compared to the temperature-sweep results shown in the main text.

## S8. Microscopic model for the interplay between superconductivity and CDW

To model the interplay between superconductivity and a charge-density wave (CDW) on a microscopic level, we make use of a mean-field theory as it was established e.g. in ref(20) to describe the coexistence of superconductivity and a phonon-mediated CDW in NbSe$_2$. This model does not incorporate any cuprate specific details such that a quantitative agreement with the experiment cannot be expected. It rather serves as a minimal model describing the basic interplay and coupling mechanism between two collective modes.

We start from the following BCS Hamiltonian coupled to light and include a CDW phonon

$$H(t) = \sum_{k,\sigma} \epsilon_k c^\dagger_{k,\sigma} c_{k,\sigma} - V \sum_{k,k'} c^\dagger_{k,\uparrow} c^\dagger_{-k,\downarrow} c_{-k',\downarrow} c_{k',\uparrow} + \sum_{q=\pm Q} \omega_q b^\dagger_q b_q$$
$$+ g \sum_{k,q=\pm Q,\sigma} g_k c^\dagger_{k+q,\sigma} c_{k,\sigma} (b_q + b^\dagger_{-q}) + \frac{1}{2} \sum_{k,\sigma} \sum_{i,j} \partial^2_{ij} \epsilon_k A_i(t) A_j(t) c^\dagger_{k,\sigma} c_{k,\sigma}$$

Hereby, $\epsilon_k = \xi_k - \epsilon_F$ is the electron band dispersion measured relative to the Fermi level $\epsilon_F$, $c^\dagger_k$ and $c_k$ the electron creation and annihilation operators and $V$ an s-wave pairing interaction. Furthermore, we consider a single phonon with wave vector $Q$, which will be responsible for creating the CDW. The corresponding phonon creation and annihilation operators are $b^\dagger_Q$ and $b_Q$ and the phonon energy is $\omega_Q$. The phonon interacts with the electrons via the electron-phonon coupling strength $g \cdot g_k$, where $g$ is the strength and $g_k$ the symmetry of the interaction. The coupling to an external gauge field $A(t)$ is obtained via minimal coupling $\epsilon_k \rightarrow \epsilon_{k-A(t)}$, where we expand in orders of $A(t)$

$$\epsilon_{k-A(t)} = \epsilon_k - \sum_i \partial_i \epsilon_k A_i(t) + \frac{1}{2} \sum_{i,j} \partial^2_{ij} \epsilon_k A_i(t) A_j(t) + \mathcal{O}(A(t)^3)$$

using $\partial_i = \partial_{k_i}$ and $\partial^2_{ij} = \partial_{k_i} \partial_{k_j}$ for shorthand notation. As the system is parity symmetric, the linear coupling in $A$ vanishes and only the diamagnetic coupling remains.

To simplify the calculation, we will consider a tight-binding dispersion on a square lattice at half-filling $\epsilon_k = -2t (\cos k_x + \cos k_y)$ such that the system has perfect particle-hole symmetry. Thus, a CDW vector $Q = (\pi, \pi)$ yields perfect nesting and the CDW is commensurate with $k + 2Q \triangleq k$, i.e. $\epsilon_{k+Q} = - \epsilon_k$ or $\epsilon_{k+2Q} = \epsilon_k$. For the symmetry of the electron-phonon coupling we assume an anisotropic s-wave with $g_k = |\cos k_x - \cos k_y|$.

We develop an effective field theory description with the partition function

$$\mathcal{Z} = \int \mathcal{D}(c^\dagger, c, b^\dagger, b) e^{-S(c^\dagger, c, b^\dagger, b)},$$

where the action $S$ in the Matsubara formalism is given by

$$S(c^\dagger, c, b^\dagger, b) = \int_0^\beta d\tau \left( \sum_{k,\sigma} c_{k,\sigma}^\dagger(\tau) \partial_\tau c_{k,\sigma}(\tau) + \sum_{q=\pm Q} b_q^\dagger(\tau) \partial_\tau b_q(\tau) + H(\tau) \right).$$

Decoupling the interaction with a Hubbard-Stratonovich transformation and rewriting the system with the four-component Nambu-spinor $\psi_k^\dagger = \left( c_{k,\uparrow}^\dagger, c_{k+Q,\uparrow}^\dagger, c_{-k,\downarrow}, c_{-(k+Q),\downarrow} \right)$, we obtain in frequency representation (see also ref(20))

$$S(\psi^\dagger, \psi, \delta\Delta, \delta D) = \beta \frac{\Delta^2}{V} + \beta \frac{D^2}{W} + \frac{1}{\beta} \sum_{i\omega_m} \delta\Delta(i\omega_m) \frac{1}{V} \delta\Delta(-i\omega_m) - \frac{1}{g^2} \frac{1}{\beta} \sum_{i\omega_m} \delta D(i\omega_m) P_0^{-1}(i\omega_m) \delta D(-i\omega_m)$$

$$- \frac{1}{\beta^2} \sum_{i\omega_m, i\omega_n} \sum_{k,k'} \psi_k^\dagger(i\omega_m) G^{-1}(k, k', i\omega_m, i\omega_n) \psi_{k'}^\dagger(i\omega_n)$$

with the bare phonon propagator

$$P_0^{-1}(i\omega_m) = -\frac{\omega_Q^2 - (i\omega_m)^2}{2\omega_Q}$$

and the BCS Green's function with Pauli matrices $\tau_i$

$$G^{-1}(k, k', i\omega_m, i\omega_n) = G_0^{-1}(k, k', i\omega_m, i\omega_n) - \Sigma(k, k', i\omega_m - i\omega_n),$$

$$G_0^{-1}(k, k', i\omega_m, i\omega_n) = \left( i\omega_m \tau_0 \otimes \tau_0 - h_k^{(0)} \right) \beta \delta_{k,k'} \delta_{i\omega_m, i\omega_n} = G_0^{-1}(k, i\omega_m) \beta \delta_{k,k'} \delta_{i\omega_m, i\omega_n},$$

$$\Sigma(k, k', i\omega_m - i\omega_n) = h_k^{(1)}(i\omega_m - i\omega_n) \delta_{k,k'}.$$

Here the cross product of Pauli matrices $\tau_i \otimes \tau_j$, $i, j = 0, \ldots 3$, lies in the 4-component SC-CDW extended Nambu space: in particular, the first matrix refers to the SC subspace, while the second one acts in the CDW subspace. Hereby, the mean-field Hamiltonian $H(t) = \sum_k \psi_k^\dagger h_k(t) \psi_k$ in Nambu basis is defined as

$$h_k(t) = h_k^{(0)} + h_k^{(1)}(t),$$

$$h_k^{(0)} = \epsilon_k \tau_3 \otimes \tau_3 - \Delta \tau_1 \otimes \tau_0 - D g_k \tau_3 \otimes \tau_1,$$

$$h_k^{(1)}(t) = \frac{1}{2} \sum_{i,j} \partial_{ij}^2 \epsilon_k A_i(t) A_j(t) \tau_3 \otimes \tau_3 - \delta\Delta(t) \tau_1 \otimes \tau_0 - g_k \delta D(t) \tau_3 \otimes \tau_1.$$

In these expressions, we introduce the superconducting order parameter $\Delta$ and CDW order parameter $D_k = D g_k$ defined as

$$\Delta = 2V \sum_k \langle c_{-k,\downarrow} | c_{k,\uparrow} \rangle = V \sum_k \frac{\Delta}{E_k} \tanh\left(\frac{\beta E_k}{2}\right),$$

$$D = W \sum_{k,\sigma} g_k \langle c^\dagger_{k,\sigma} c_{k+Q,\sigma} \rangle = W \sum_k g_k^2 \frac{D}{E_k} \tanh\left(\frac{\beta E_k}{2}\right),$$

with $W = 4g^2/\omega_Q$, where we allow amplitude fluctuations $\Delta(t) = \Delta + \delta\Delta(t)$ and $D(t) = D + \delta D(t)$. Please note, as the system has perfect particle-hole symmetry, phase fluctuations can be neglected as any coupling to long-range Coulomb interaction is identical zero[5]. A diagonalization of the Hamiltonian yields the quasiparticle energy $E_k = \sqrt{\epsilon_k^2 + \Delta^2 + |D_k|^2}$.

Next, we integrate out the fermions using $\int \mathcal{D}(\psi^\dagger,\psi) e^{-\psi^\dagger X \psi} = e^{\text{tr}(\ln X)}$, where the trace includes the momentum and Matsubara sum. The logarithm can be expanded for small $\Sigma$ via $\text{tr}\ln(-G^{-1}) = \text{tr}\ln(-G_0^{-1}) - \text{tr}\sum_{n=1}^{\infty} \frac{(G_0 \Sigma)^n}{n}$ and the action is split into a mean-field part and a fluctuation part

$$S(\delta\Delta, \delta D) = S_{\text{MF}} + S_{\text{FL}}(\delta\Delta, \delta D)$$

with

$$S_{\text{MF}} = \beta \frac{\Delta^2}{V} + \beta \frac{D^2}{W} - \text{tr}\ln(-G_0^{-1}),$$

$$S_{\text{FL}}(\delta\Delta, \delta D) = \frac{1}{\beta} \sum_{i\omega_m} \delta\Delta(i\omega_m) \frac{1}{V} \delta\Delta(-i\omega_m) - \frac{1}{g^2} \frac{1}{\beta} \sum_{i\omega_m} \delta D(i\omega_m) P_0^{-1}(i\omega_m) \delta D(-i\omega_m) + \text{tr} \sum_{n=1}^{\infty} \frac{(G_0 \Sigma)^n}{n}.$$

As we are interested in the THG signal $j^{(3)} = \delta S/\delta A \propto A^3$, the terms in the action with power of $A^4$ are relevant. Thus, we consider the second order term in the sum of the trace $\frac{1}{2}\text{tr}(G_0 \Sigma G_0 \Sigma)$. Neglecting the bare response, it follows for the fourth order action

$$S^{(4)}(\phi) = \frac{1}{2} \frac{1}{\beta} \sum_{i\omega_m} [\phi^\top(-i\omega_m) M(i\omega_m) \phi(i\omega_m) + \phi^\top(-i\omega_m) b(i\omega_m) + b^\top(-i\omega_m) \phi(i\omega_m)]$$

with

$$\phi^\top(-i\omega_m) = \begin{pmatrix} \delta\Delta(-i\omega_m) & \delta D(-i\omega_m) \end{pmatrix},$$

$$M(i\omega_m) = \begin{pmatrix} H^{-1}(i\omega_m) & \chi_{\Delta D}(i\omega_m) \\ \chi_{D\Delta}(i\omega_m) & -\frac{1}{g^2} P^{-1}(i\omega_m) \end{pmatrix},$$

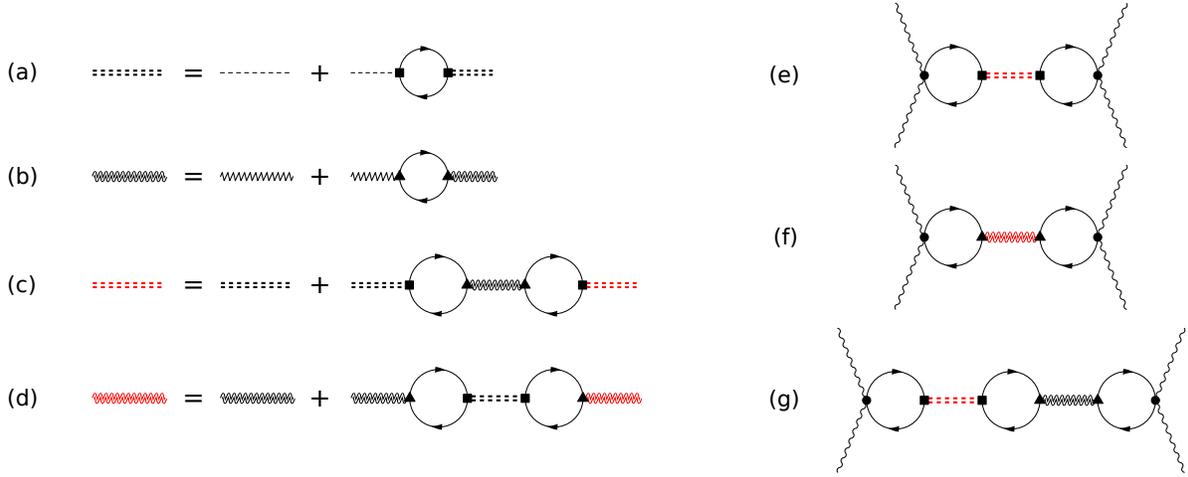

**Figure S6 Diagrammatic representation of the terms occuring in the effective action. (a)** Higgs propagator defined as the renormalization of the pairing interaction. **(b)** Renormalized phonon propagator due to electron-phonon interaction. **(c)** Renormalized Higgs propagator due to interaction with CDW **(d)** Renormalized phonon propagator due to interaction with Higgs **(e),(f),(g)** Final terms in fourth order action, i.e. **(e)** Higgs, **(f)** CDW, **(g)** mixed contribution.

$$b(i\omega_m) = \begin{pmatrix} -\sum_{ij} \chi^{ij}_{\Delta A^2}(i\omega_m) A^2_{ij}(i\omega_m) \\ -\sum_{ij} \chi^{ij}_{DA^2}(i\omega_m) A^2_{ij}(i\omega_m) \end{pmatrix}.$$

In these expressions, we defined the Higgs progagator as

$$H^{-1}(i\omega_m) = \frac{2}{V} + \chi_{\Delta\Delta}(i\omega_m) = 2\sum_k \frac{4\Delta^2 - (i\omega_m)^2}{E_k(4E_k^2 - (i\omega_m)^2)} \tanh(\beta E_k/2)$$

which can be understood as an RPA series renormalization of the pairing interaction as shown diagrammatically in Fig. S6a. Analogously, the renormalized phonon propagator is defined as shown in Fig. S6b. It reads

$$P^{-1}(i\omega_m) = P_0^{-1}(i\omega_m) - g^2 \chi_{DD}(i\omega_m) = -\frac{\Omega_Q^2 - (i\omega_m)^2}{2\omega_Q}$$

with

$$\Omega_Q^2 = \omega_Q^2 + 2\omega_Q g^2 \chi_{DD}(i\omega_m) = \omega_Q^2 W \sum_k g_k^2 \frac{4D_k^2 - (i\omega_m)^2}{E_k(4E_k^2 - (i\omega_m)^2)} \tanh(\beta E_k/2).$$

Hereby, the susceptibilities are defined as

$$X_{\alpha\beta\gamma\delta}(\mathbf{k}, i\omega_n) = \frac{1}{\beta} \sum_{i\omega_n} \text{tr}\left[G_0(\mathbf{k}, i\omega_n)\, \tau_\alpha \otimes \tau_\beta\, G_0(\mathbf{k}, i\omega_m + i\omega_n)\, \tau_\gamma \otimes \tau_\delta\right],$$

$$\chi_{\Delta\Delta}(i\omega_m) = \sum_{\mathbf{k}} X_{1010} = -8 \sum_{\mathbf{k}} \frac{D_\mathbf{k}^2 + \epsilon_\mathbf{k}^2}{E_\mathbf{k}(4E_\mathbf{k}^2 - (i\omega_m)^2)} \tanh(\beta E_\mathbf{k}/2),$$

$$\chi_{\Delta D}(i\omega_m) = \sum_{\mathbf{k}} g_\mathbf{k} X_{1031} = 8 \sum_{\mathbf{k}} g_\mathbf{k} \frac{\Delta D_\mathbf{k}}{E_\mathbf{k}(4E_\mathbf{k}^2 - (i\omega_m)^2)} \tanh(\beta E_\mathbf{k}/2),$$

$$\chi_{\Delta A^2}^{ij}(i\omega_m) = \sum_{\mathbf{k}} \frac{1}{2} \partial_{ij}^2 \epsilon_\mathbf{k} X_{1033} = -4 \sum_{\mathbf{k}} \partial_{ij}^2 \epsilon_\mathbf{k} \frac{\epsilon_\mathbf{k} \Delta}{E_\mathbf{k}(4E_\mathbf{k}^2 - (i\omega_m)^2)} \tanh(\beta E_\mathbf{k}/2),$$

$$\chi_{DD}(i\omega_m) = \sum_{\mathbf{k}} g_\mathbf{k}^2 X_{3131} = -8 \sum_{\mathbf{k}} g_\mathbf{k}^2 \frac{\Delta^2 + \epsilon_\mathbf{k}^2}{E_\mathbf{k}(4E_\mathbf{k}^2 - (i\omega_m)^2)} \tanh(\beta E_\mathbf{k}/2),$$

$$\chi_{DA^2}^{ij}(i\omega_m) = \sum_{\mathbf{k}} g_\mathbf{k} \frac{1}{2} \partial_{ij}^2 \epsilon_\mathbf{k} X_{3133} = -4 \sum_{\mathbf{k}} g_\mathbf{k} \partial_{ij}^2 \epsilon_\mathbf{k} \frac{\epsilon_\mathbf{k} D_\mathbf{k}}{E_\mathbf{k}(4E_\mathbf{k}^2 - (i\omega_m)^2)} \tanh(\beta E_\mathbf{k}/2),$$

with the symmetry property $X_{\gamma\delta\alpha\beta}(\mathbf{k}, i\omega_m) = X_{\alpha\beta\gamma\delta}(\mathbf{k}, -i\omega_m)$. Finally, we integrate out the amplitude fluctuations obtaining $S^{(4)} \propto b^\mathsf{T} M^{-1} b$ with the inverse of $M$

$$M^{-1}(i\omega_m) = \begin{pmatrix} \tilde{H}(i\omega_m) & g^2 \chi_{\Delta D}(i\omega_m) P(i\omega_m) \tilde{H}(i\omega_m) \\ g^2 \chi_{D\Delta}(i\omega_m) \tilde{P}(i\omega_m) H(i\omega_m) & -g^2 \tilde{P}(i\omega_m) \end{pmatrix}$$

where we identify the renormalized Higgs and phonon propagator

$$\tilde{H}(i\omega_m) = \frac{1}{H^{-1}(i\omega_m) + g^2 \chi_{\Delta D}(i\omega_m) \chi_{D\Delta}(i\omega_m) P(i\omega_m)}$$

$$\tilde{P}(i\omega_m) = \frac{1}{P^{-1}(i\omega_m) + g^2 \chi_{\Delta D}(i\omega_m) \chi_{D\Delta}(i\omega_m) H(i\omega_m)}$$

which can be again understood as an RPA renormalization as shown in Fig. S6c-d. For the further analysis, we assume linear polarized light in $x$-direction, i.e.

$$A(t) = \begin{pmatrix} A_0 \cos(\Omega t) \\ 0 \end{pmatrix},$$

$$A^2(\omega) = A_0^2 \left( \delta(\omega) + \frac{1}{2}\delta(\omega - 2\Omega) + \frac{1}{2}\delta(\omega + 2\Omega) \right).$$

As a result, only the derivative $\partial_{xx}^2 \epsilon_\mathbf{k}$ remains in the respective susceptibilities such that we neglect the polarization indices in the following. With this, the fourth order action finally reads after analytic continuation $i\omega_m \to \omega + i0^+$

$$S^{(4)} = \frac{1}{2} \int d\omega\, K^{(4)}(\omega) A^2(\omega) A^2(-\omega)$$

where the kernel is given by

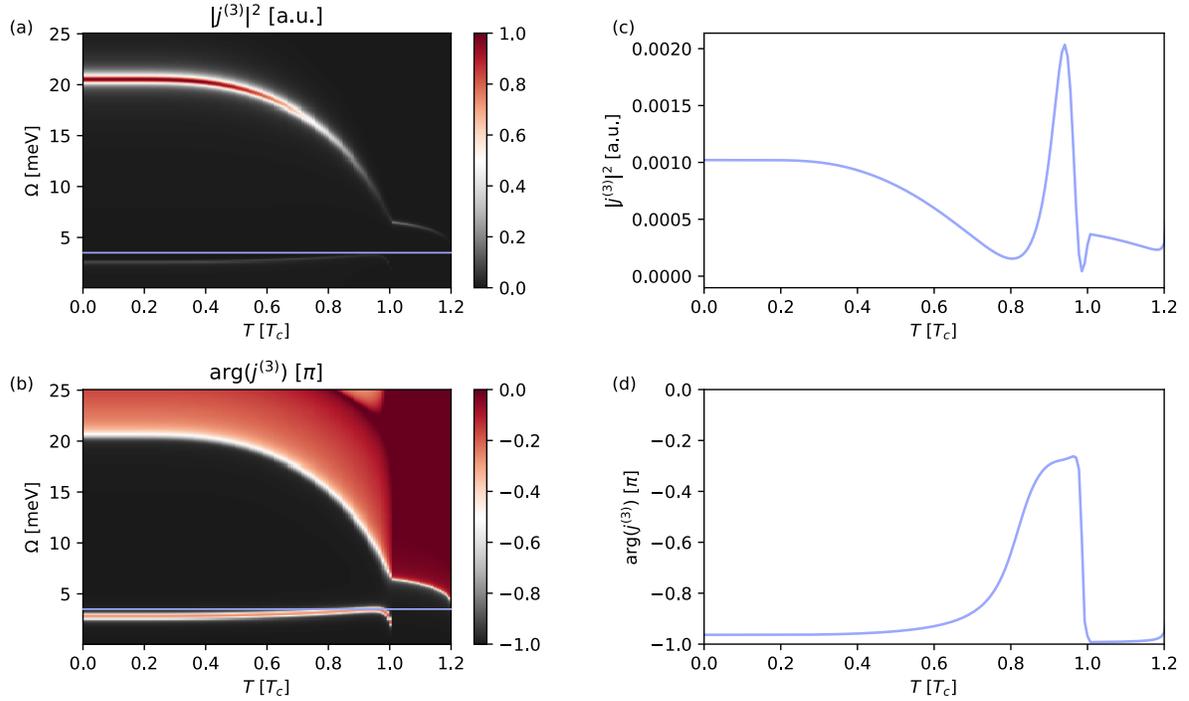

**Figure S7 THG signal as function of frequency and temperature. (a)** Intensity and **(b)** phase of THG signal. **(c)** THG intensity and **(d)** THG phase as function of temperature for fixed driving frequency corresponding to the horizontal lines.

$$K^{(4)}(\omega) = -\chi_{A^2\Delta}(\omega)^2 \tilde{H}(\omega) - g^2 \chi_{A^2D}(\omega)^2 \tilde{P}(\omega) + 2g^2 \chi_{A^2\Delta}(\omega)\chi_{A^2D}(\omega)\chi_{\Delta D}(\omega)\tilde{H}(\omega)P(\omega).$$

The three contributions are related to the Higgs excitation, CDW excitation and a mixed term which are shown in Fig. S6e-g. As a result, the induced THG current is directly proportional to the fourth order kernel

$$j^{(3)}(3\Omega) = -\left.\frac{\delta S^{(4)}}{\delta A_\alpha(-\omega)}\right|_{3\Omega} \propto K^{(4)}(2\Omega)$$

We evaluate the three diagrams using the parameters $\Delta = 20$ meV, $D = 10$ meV, $\omega_0 = 20$ meV, $t = 100$ meV, and a broadening $\Omega + i\eta$ with $\eta = 0.1$ meV on a 2d momentum grid with $2000 \times 2000$ points for varying driving frequency and temperature. Hereby, the superconducting gap $\Delta(T)$ and CDW gap $D(T)$ is self-consistently evaluated for each temperature by iteratively solving the two gap equations simultaneously. It is important to emphasize that this model does not quantitatively describe the experiment on cuprates but serves as a general proof of principle for the interplay between the Higgs mode and a CDW fluctuation. As one can see in Fig. S7, two resonances appear in the THG spectrum corresponding to the renormalized Higgs mode at ~ 20 meV and at the renormalized phonon frequency at ~ 3 meV at $T = 0$. The temperature dependence roughly follows the temperature dependence of the order parameter. Each resonance is accompanied by a positive phase change of roughly $\pi$ slightly reduced due to the broadening. The phase signature is more clear than the peak in the amplitude for the lower mode. In addition, the interference between the two modes is visible by an antiresonance characterized by a negative phase change of roughly $\pi$ close to the lower mode. As one can see, a horizontal cut in the 2d plot near the lower mode shows the negative phase change as observed in the experiment. The exact shape of the horizontal cuts depend on the details of the model, yet the antiresonance behavior with a negative phase change is always visible as long as the driving frequency matches the energy of antiresonance for some temperature.

## S9. Evidence for anti-resonance above $T_c$

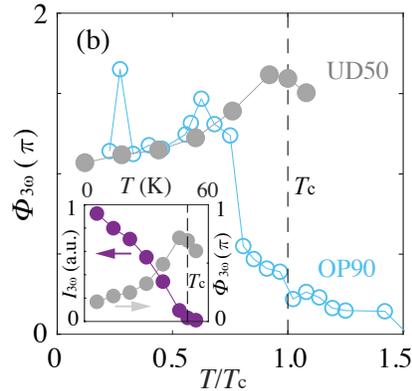

**Figure S8 THG from bilayer cuprates** THG phase response of an underdoped $DyBa_2Cu_2O_{7-x}$ (UD50) measured with 0.5 THz driving frequency, superposed on top of the THG phase response of an optimally doped $DyBa_2Cu_2O_{7-x}$(OP90) measured with 0.7 THz driving frequency as reported in ref(3). Inset shows the temperature dependence of THG intensity and phase of the UD50 sample. $\Phi_{3\omega}$ cannot be reliably extracted beyond $T_c$ as the THG signal becomes very low there, due in part to the use of a table-top THz source for this measurement. It can be seen from the main figure that an anti-resonance seems to appear just above $T_c$.

We measured the THG response of an underdoped $DyBa_2Cu_2O_{7-x}$(UD50) film with 0.5 THz driving frequency. A positive evolution of the THG phase is observed below $T_c$ with the start of a downturn right at $T_c$, indicating a potential anti-resonance above $T_c$. We compare the results of this UD50 sample to an optimally doped $DyBa_2Cu_2O_{7-x}$(OP90) sample measured previously with 0.7 THz driving frequency. The anti-resonance in that case takes place significantly below $T_c$. These observations are consistent with the coupled oscillators model/Fano resonance model, which predicts that the anti-resonance moves to a lower temperature when using a higher driving frequency.

**Reference**

1. Green, B. *et al.*, High-Field High-Repetition-Rate Sources for the Coherent THz Control of Matter. *Sci. Rep.* **6**, 22256 (2016).

2. Kovalev, S. *et al.* Probing ultra-fast processes with high dynamic range at 4th-generation light sources: Arrival time and intensity binning at unprecedented repetition rates. *Structural Dynamics* **4**, 024301 (2017).

3. Chu, H. *et al.* Phase-resolved Higgs response in superconducting cuprates. *Nat. Commun.* **11** 1793 (2020).

4. Katsumi, K. *et al.* Higgs Mode in the *d*-Wave Superconductor $Bi_2Sr_2CaCu_2O_{8+x}$ Driven by an Intense Terahertz Pulse. *Phys. Rev. Lett.* **120**, 117001 (2018).

5. Matsunaga, R. *et al.* Light-induced collective pseudospin precession resonating with Higgs mode in a superconductor. *Science* **345**, 1145-1149 (2014).

6. Tsuji, N. & Aoki, H. Theory of Anderson pseudospin resonance with Higgs mode in superconductors. *Phys. Rev. B* **92**, 064508 (2015).

7. Tsuji, N. *et al.* Nonlinear light–Higgs coupling in superconductors beyond BCS: Effects of the retarded phonon-mediated interaction. *Phys. Rev. B* **94**, 224519 (2016).

8. Schwarz, L. & Manske, D. Theory of driven Higgs oscillations and third-harmonic generation in unconventional superconductors. *Phys. Rev. B* **101**, 184519 (2020).

9. Cea, T. *et al.* Nonlinear optical effects and third-harmonic generation in superconductors: Cooper pairs versus Higgs mode contribution. *Phys. Rev. B* **93**, 180507(R) (2016).

10. Seibold, G. *et al.* Third harmonic generation from collective modes in disordered superconductors. *Phys. Rev. B* **103**, 014512 (2021).

11. Gabriele, F. *et al.* Non-linear Terahertz driving of plasma waves in layered cuprates. *Nat. Commun.* **12**, 752 (2021).

12. Tsuji, N. & Nomura, Y. Higgs-mode resonance in third harmonic generation in NbN superconductors: Multiband electron-phonon coupling, impurity scattering, and polarization-angle dependence. *Phys. Rev. Res.* **2**, 043029 (2020).

13. Silaev, M. Nonlinear electromagnetic response and Higgs-mode excitation in BCS superconductors with impurities. *Phys. Rev. B* **99**, 224511 (2019).

14. Murotani, Y. & Shimano, R. Nonlinear optical response of collective modes in multiband superconductors assisted by nonmagnetic impurities. *Phys. Rev. B* **99**, 224510 (2019).

15. Haenel, R. *et al.* Time-resolved optical conductivity and Higgs oscillations in two-band dirty superconductors. *Phys. Rev. B* **104**, 134504 (2021).

16. Puviani, M. *et al.* Calculation of an enhanced A1g symmetry mode induced by Higgs oscillations in the Raman spectrum of high-temperature cuprate superconductors. Preprint available at https://arxiv.org/abs/2012.01922v2 to appear in *Phys. Rev. Lett.*

17. Cea. T. *et al.* Polarization dependence of the third-harmonic generation in multiband superconductors. *Phys. Rev. B* **97**, 094516 (2018).

18. Torchinsky, D. *et al.* Structural Distortion-Induced Magnetoelastic Locking in $Sr_2IrO_4$ Revealed through Nonlinear Optical Harmonic Generation. *Phys. Rev. Lett.* **114**, 096404 (2015).

19. Iizawa, M. *et al.* The quantum and classical Fano parameter q. *Phys. Scr.* **96** 055401 (2021).

20. Cea, T. & Benfatto, L. Nature and Raman signatures of the Higgs amplitude mode in the coexisting superconducting and charge-density-wave state. *Phys. Rev. B* **90**, 224515 (2014).